\documentclass{emulateapj}
\usepackage{graphicx}
\begin{document}
\newcommand{\vdag}{(v)^\dagger}
\newcommand{\av}{$A_{V}$}
\newcommand{\mav}{A_{V}}
\newcommand{\degrees} {$^\circ$}
\newcommand{\kms}{km\ s$^{-1}$}
\newcommand{\cc}{cm$^{-3}$}
\newcommand{\htwo}{${\rm H_2}$}
\newcommand{\mhtwo}{{\rm H_2} }
\newcommand{\cotwoone}{$^{12}$CO~(2-1)}
\newcommand{\co}{$^{12}$CO}
\newcommand{\thco}{$^{13}$CO}
\newcommand{\mco}{ {\rm ^{12}CO}}
\newcommand{\mthco} { {\rm ^{13}CO}}
\newcommand{\wthco}{$W(\mthco)$}
\newcommand{\nthco}{$N(\mthco)$}
\newcommand{\myemail}{agoodman@cfa.harvard.edu}
\def \simless {\mathbin{\lower 3pt\hbox{$\rlap{\raise 6pt
              \hbox{$\char'074$}}\mathchar"7218$}}}
\def \simgreat {\mathbin{\lower 3pt\hbox{$\rlap{\raise 6pt
              \hbox{$\char'076$}}\mathchar"7218$}}}

\slugcomment{Accepted in ApJ}

\shorttitle{Column Density Distribution}
\shortauthors{A.~A. Goodman, J.~E. Pineda \& S.~L. Schnee}

\title{The ``True" Column Density Distribution in Star-Forming Molecular Clouds}

\author{Alyssa A. Goodman}
\affil{Harvard-Smithsonian Center for Astrophysics, 60 Garden Street, MS-42, Cambridge, MA 02138, USA}
\email{agoodman@cfa.harvard.edu}
\author{Jaime E. Pineda}
\affil{Harvard-Smithsonian Center for Astrophysics, 60 Garden Street, MS-10, Cambridge, MA 02138, USA}
\email{jpineda@cfa.harvard.edu}
\and
\author{Scott L. Schnee}
\affil{Harvard-Smithsonian Center for Astrophysics, 60 Garden Street, MS-10, Cambridge, MA 02138}
\affil{Division of Physics, Mathematics and Astronomy, California Institute of Technology, 770 South Wilson Avenue, Pasadena, CA 91125, USA}
\email{schnee@phobos.caltech.edu}

\begin{abstract}
We use the COMPLETE Survey's observations of the Perseus star-forming region to assess and intercompare three methods for measuring column density in molecular clouds: near-infrared extinction mapping; thermal emission
mapping in the far-IR; and mapping the intensity of CO isotopologues. Overall, the structures shown by all three tracers are morphologically similar, but important differences exist amongst the tracers.  We find that the dust-based measures (near-IR extinction and thermal emission) give similar, {\it log-normal}, distributions for the full ($\sim 20$ pc-scale) Perseus region, once careful calibration corrections are made.   We also compare dust- and gas-based column density distributions for physically-meaningful sub-regions of Perseus, and we find significant variations in the distributions for those (smaller, $\sim$few pc-scale) regions. Even though we have used \co\ data to estimate excitation temperatures, and we have corrected for opacity, the \thco\ maps seem unable to give column distributions that consistently resemble those from dust measures.  We have edited out the effects of the shell around the B-star HD 278942 from the column-density distribution comparisons. In that shell's interior and in the parts where it overlaps the molecular cloud, there appears to be a dearth of \thco, which is likely due either to \thco\ not yet having had time to form in this young structure, and/or destruction of \thco\ in the molecular cloud by the HD 278942's wind and/or radiation.  We conclude that the use of either dust or gas measures of column density without extreme attention to calibration (e.g. of thermal emission zero-levels) and artifacts (e.g. the shell) is more perilous than even experts might normally admit.  And, the use of \thco\ data to trace total column density in detail, even after proper calibration, is unavoidably limited in utility due to threshold, depletion, and opacity effects.  If one's main aim is to map column density (rather than temperature or kinematics), then dust extinction seems the best probe, up to a limiting extinction caused by a dearth of sufficient background sources.  Linear fits amongst all three tracers' estimates of column density are given, allowing us to quantify the inherent uncertainties in using one tracer, in comparison with the others.  
\end{abstract}

\keywords{dust, extinction --- ISM:abundances --- ISM:molecules --- 
ISM:individual (Perseus molecular complex)}

\section{Introduction}

In this paper, it is our goal to use data from the COMPLETE Survey of
Star-Forming Regions\footnote{All of the data from the COMPLETE
(\textbf{CO}ordinated \textbf{M}olecular \textbf{P}robe \textbf{L}ine
\textbf{E}xtinction \textbf{T}hermal \textbf{E}mission) Survey are
available online at http://www.cfa.harvard.edu/COMPLETE.} to assess
and intercompare three methods for measuring column density in
molecular clouds: near-infrared extinction mapping; thermal emission
mapping in the far-IR; and mapping the intensity of CO isotopologues.  We
wish we could offer a snapshot of {\it volume} density, as numerical
modelers can, but volume density cannot be mapped from our vantage
point on Earth without making very model-dependent assumptions.  We
discuss the position-position-{\it velocity} distribution of material in
other papers \citep[e.g.][]{Erik:2007, dendro}: here we focus solely on the column density
distribution.  Our aim is to offer insight into the biases of
particular techniques, and to provide the best estimates to date of ``true"
column density distributions in molecular clouds suitable for
comparison with current and future numerical simulations.

We focus our discussion on the Perseus molecular cloud complex, which covers nearly 10
square degrees on the sky \citep[$\sim200$~pc$^2$ at 250 pc;
][]{Enoch06,Hirota:2008}. A companion paper to this one \citep{Pineda07} includes
an historical perspective on the use of dust extinction and emission
and of CO isotopologue mapping to study column density in molecular clouds,
and it offers an in-depth look at abundance variations in the cloud.

\section{Data: Three Maps of Perseus}
\label{data}
In the past, three principal methods have been used to chart column
density in molecular clouds: (1) {\bf extinction} mapping, using either
star counting 
\citep[e.g.][]{barnard,cern84,cern85}
or color-excess measurements \citep[e.g.][]{Lada:IC5146}; (2)
{\bf dust-emission} mapping, at far-IR through mm--wavelengths
\citep[e.g.][]{Schlegel98}; and (3) mapping {\bf integrated intensity of
molecular-line emission}, usually \co\ or \thco\ on large ($>1$~pc)
scales \citep[e.g.][]{Padoan:Bell}.  As noted above, there is a long
history of the application of these techniques, and we analyze that
history in \citet{Pineda07}.

All three of these methods have now been applied, as part of COMPLETE,
to the large ($\sim 10$~deg$^2$) swath of sky in
Perseus\footnote{Ophiuchus and Serpens have also been observed
similarly in the COMPLETE and c2d Surveys, but this short paper
focuses on the most studied maps to-date, which are of Perseus} which
has also been surveyed by the Spitzer Space Telescope under the
``Cores-to-Disks" \citep[c2d;][]{Evans:C2D} Legacy
Program\footnote{\url{http://peggysue.as.utexas.edu/SIRTF/}}.  The
Perseus maps upon which the analysis in this paper rests can be found
in \cite{Schnee05} and \citet{Ridge06a,Ridge06b}.

\subsection{Extinction Mapping}
\label {data-extinction}
The extinction map used here is presented in \cite{Ridge06a}.  To
create the map, COMPLETE collaborators Jo\~ao Alves and Marco
Lombardi, applied their ``NICER" (Near-Infrared Color Excess method
Revisited) method to 2MASS near-infrared maps of
Perseus\footnote{2MASS data used are from the ``Two-Micron All-Sky
Survey, Final Release" (See
http://www.ipac.caltech.edu/2mass/releases/allsky/doc/explsup.html)}.

\begin{figure*}
\centering
\epsscale{1.18}
\plotone{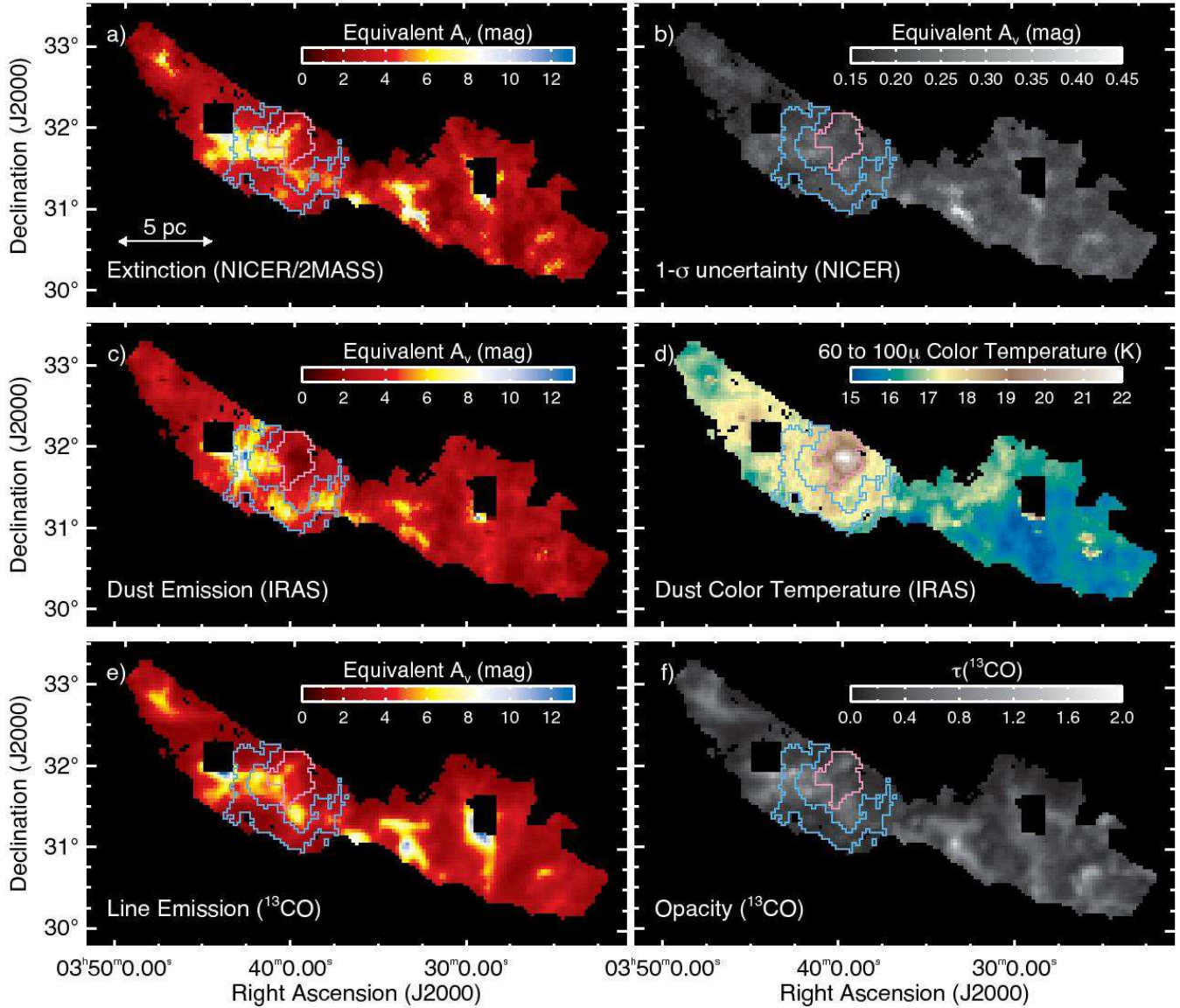}
\caption{
Maps in the left column show column density as traced by: Extinction (a); Dust Emission (c); and Gas Emission (e).  Figure 1b shows the small uncertainties associated with the (NICER) extinction map in Figure 1a.  Dust color temperatures based on 60 and 100 \micron\ data are shown in Figure 1d.  Note the warm dust associated with the shell around HD278942.  The opacity of \thco, which is correlated with column density, is shown in Figure 1f.  In all panels, only data points with detections of \thco\ with S/N$>= 5$ are shown, and the resolution is 5\arcmin.  The IC348 and NGC1333 regions have been excised from the data sets, because all three techniques are biased in cluster regions.  The single pink contour surrounds the apparently heated material around HD278942, and the single light blue contour outlined the ``overdense" area presumably created by HD278942's shell, showing the dearth of \thco\ indicated most clearly in Figure~\ref{f3}. {\bf The authors request that this figure be scaled to fill the full width of the page.}
\label{f1}}
\end{figure*}

As described in \cite{Lombardi01}, the NICER algorithm calculates extinction values
based on the difference between the
\textit{observed} average near-IR color of stars within a sampling box
to the stars' \textit{intrinsic} average color. As the number of
stars in a sampling box is reduced, the uncertainty in the extinction
calculated by NICER rises.  The details of the tradeoffs between
high-resolution/high-uncertainty and low-resolution/low-uncertainty
mapping with NICER is discussed in detail by \citet{lom06}.
The NICER/2MASS map used here is made with 5\arcmin\ resolution on a 
2.5\arcmin\ grid, a
common resolution to which all data presented in this paper have been
smoothed.  As shown in Figure \ref{f1}a, we can measure extinctions from $\mav=0$ to
10 mag with typical reliability better than 0.25~mag (Figure
\ref{f1}b), but it remains true that the highest uncertainties ($\sim0.4$~mag) 
are in the highest extinction regions ($\mav \simgreat 10$~mag).

Even though 5\arcmin\ resolution sounds coarse (especially
compared with the 46\arcsec\ intinsic resolution of the COMPLETE
spectral-line maps), it is important to appreciate that earlier,
optically-based, extinction mapping methods could never come close to
this kind of resolution in high column-density regions.  In the
$V$-band, for example, even modern star-counting techniques typically
fail (as zero stars are present in a counting box) at $\mav \simgreat 6$ mag
\citep[e.g.][]{Cambresy99}.  The material with $5\simless \mav \simless 10$ mag is
exactly the material, when observed with the kind of
few-tenths-of-a-pc resolution we have here, that appears most actively
engaged in forming stars in Perseus \citep{Kirk06,Joergensen06}, and
it is nearly completely opaque in optical wide-field surveys.

We choose, at this point in time, to use the 2MASS data for our extinction maps, rather than the Spitzer IRAC c2d data for two reasons.  First, the 2MASS data cover the full region available in COMPLETE's molecular line maps, while the IRAC data only span ``most" of the region.  Second, and more importantly, work is still ongoing (Huard, private communication) on the calibration of IRAC-based extinction maps and on the determination an agreed-upon mid-IR extinction law.  In the future, the best extinction maps will likely be constructed from photometric observations spanning the NIR and MIR range.  Since the absorption by dust becomes progressively less severe at longer wavelengths, the MIR data will be especially useful for mapping out the structure of very high column density gas.

\subsection{Far-Infrared Emission Mapping}
\label{data-far-IR}

The far infrared data used here are taken from the IRIS (Improved
Reprocessing of the IRAS Survey) database \citep{Miville05}.  IRIS
represents a substantial improvement over the earlier ISSA release of the
IRAS survey, in that it has better zodiacal light subtraction, destriping,
calibration and zero level determinations.  The
importance of finding and using the proper zero level when deriving
column density maps where fluxes are not much higher than the noise is
explained in \citet{Arce01}.  For IRAS-based observations of clouds
like Perseus, the FIR signal is typically only much greater than the
noise above $\mav \sim 2$~mag.\footnote{\cite{Schnee2007} have recently used the long-wavelength data from Spitzer to produce higher-resolution thermal-emission maps of column density. As we are forced to smooth to 5\arcmin\ resolution in this paper (to match the near-IR extinction maps) comparison of IRAS-based maps with the newer, Spitzer-based, maps is presented and discussed in \citealt{Schnee2007}, but not here.}

Using the IRIS flux maps at 60 and 100 \micron, we derive dust
temperature and FIR optical depth maps, as explained in
\cite{Schnee05}.   We use the 2MASS/NICER column density map described
in \S 2.1 to calibrate the conversion between FIR optical depth
and NIR extinction.  Thus, we fix the overall column
density scale with absorption measurements, and then adjust the dust
emission properties, given derived color temperatures, to
minimize differences beetween absorption and emission probes of dust
column. (Note that this procedure assures that the NIR (extinction) vs. FIR (emission) column desnity plots shown in Figure~~\ref{f3} have slope and intercept close to one and zero, respectively.) 
By comparing Figure \ref{f1}c, which shows the re-calibrated,
IRAS/IRIS-derived ``equivalent \av" column density map, to the
2MASS/NICER map in Figure \ref{f1}a, one can see that the point-to-point agreement is good,
but far from perfect (see \S \ref{results}).  Figure \ref{f1}d shows the derived
color-temperature map (from \citealt{Schnee05}) that is used in creating
Figure \ref{f1}c.

Note, as with the choice made for extinction maps, that we have not chosen to use Spitzer maps of thermal emission in the present study.  Recently, \citet{Schnee2007} have presented a full analysis of the long wavelength Spitzer MIPS maps of Perseus, in concert with the IRAS-based data used here.  The Spitzer map coverage is smaller than the IRAS (all-sky) coverage, and at the resolution of the present study, the results of \citet{Schnee2007} show that the subtle differences between extinctions determined using 60 and 100 microns, and combinations at longer wavelengths are not critical to our analysis here (see histograms in Figure 9 of \citealt{Schnee2007}).

\subsection{Molecular-Line Mapping}
\label{data-molecular-line}
The implicit assumption behind mapping column density using molecular
line emission is that by integrating emission over all velocities, one
can trace out all of the molecular gas along any particular line of
sight.  This assumption fails when either: 1) a molecular line is only
excited under special physical conditions; 2) emission from a
particular line becomes optically thick; and/or 3) the species used in
an investigation does not have a constant abundance relative to
molecular hydrogen.  In reality, any one and often all three of these
are true \citep[see][]{Pineda07}. But still, by making 
a prudent choice of molecular tracer, one can minimize these complicating conditions and use line maps to trace column density.

In the past, researchers \citep{bc86,Langer89} have argued that \thco, which is excited
above volume densities $\sim 1000$ cm$^{-3}$ (column densities
$\mav \sim 1$~mag in nearby molecular
clouds) remains optically thin throughout ``most" of a molecular
cloud's volume, and also remains of relatively constant abundance,
except in very cold ($T<15~\rm{K}$) dense ($n(\mhtwo) >
5000$~\cc)  regions where carbon-bearing species are heavily depleted
\citep{caselli99}.  Therefore,
\thco\ maps are used here as the most relevant molecular line tracer
to compare with extinction maps and IRIS-based dust emission maps,
which are also sensitive to material at or above about 1~mag.

The full \thco\ map presented here was made for the COMPLETE Survey at
the Five College Radio Astronomy Observatory, and it is described in
detail in \cite{Ridge06a}.  The full-resolution map was made in
on-the-fly (OTF) mode with the 32-element SEQUOIA array, and contains
nearly 200,000 independent 46\arcsec\ pixels.  As noted above, we have
smoothed the map to 5\arcmin\ resolution in order to match the
column density maps described in \S\ref{data-extinction} and
\S\ref{data-far-IR} above.  After this smoothing, the average baseline
rms in each of the resulting pixels is 0.027~K, and there
are just over 3000 pixels (in all the maps in Figure \ref{f1}).  All points 
with signal-to-noise ratios (based on peak/rms temperature) below 5 in the smoothed map have been
excised as less-than-reliable \thco\ detections.

One of the most common, but not necessarily most accurate, ways to estimate molecular gas column density, $N(\mhtwo)$, from \thco\ emission comes from: adjusting the velocity-integrated intensity of \thco,
\wthco~$=\int {T_A (\mthco) dv}$, for telescope efficiency; assuming a uniform excitation temperature; and multiplying by a conversion factor that accounts for the ratio of \htwo\ to \thco\ 
\citep[e.g.][and references therein]{bc86}.  

To make a more accurate calculation of $N(\mthco)$ one can employ measured kinetic temperatures and
optical depths from \co\ observations \citep[see][]{Langer89}. Because COMPLETE includes
\co\ as well as \thco\ maps, we can use the observed
brightness temperature of \co\ to estimate the excitation temperature
(assuming that \co\ is optically thick), and then calculate a column density map 
assuming that the levels are populated following a Boltzmann distribution.  All maps and graphs relating to \thco\ column density distributions in this paper, unless explicitly labeled as ``{\it W}(\thco )", are made in this way \citep[see][]{Ridge06a}.  We also assume a constant
abundance, of $3.98\times 10^5$, for \htwo\ relative to \thco, 
even though some important variations in that ratio do exist, and are studied in
detail in \cite{Pineda07}.  

Figure \ref{f1}f shows the opacities derived from
the same set of calculations that yields Figure \ref{f1}e (the column density
map). The opacities, many of which significantly exceed unity, are
well-correlated with the column density. Once $\tau \gg 1$, any spectral line is
no longer a very faithful column density tracer, even when we work arduously
to correct for opacity. 

In order to convert molecular column density, $N$, to the ``equivalent \av" units used in this paper, we use the procedure outlined in \citet[][]{Pineda07}, which gives:
\begin{equation}\label{eq:n13-lte}
\mav(\mthco)= 4.24\times 10^{-16} N(\mthco)+1.67~.
\end{equation}
To facilitate comparison of the present analysis with previous work,  \citet{Pineda07} assume a constant ratio of reddening to extinction, $R_V=\mav/E(B-V)=3.1$, equal to the measured average value measured for the Galaxy \citep{bohlin:1978}.  We note, though, that $R_V$ can have values up to $\sim 6$ \citep{Draine:Review, Goodman95}, especially in high-density regions.  Lastly, we point  out that the coefficients in eq.~(\ref{eq:n13-lte}) apply to {\it Perseus as a whole}, even though we measure them to vary by as much as $\pm 30\%$  from region to region within Perseus \citep{Pineda07}.  We discuss the significance of these regional variations in \S \ref{regvar} below.

In the bottom panel of Figure \ref{f2}, we show both the \wthco-based and the \nthco-based histograms of column density.\footnote{Throughout the paper, we quote column density in units of \av, but there are assumptions used to reach such units, which are explained in \S\ref{data}.} Notice that the \wthco\ distribution appears to underestimate the amount of material at low column density (where temperature is typically higher) and at very high column density (where $\tau$ is high).  Even after accounting for temperature and opacity variations though, the $N(\mthco)$ distribution is still dangerous to interpret too literally, because at low \av\ ($<$3~mag) the gas is sub-thermally excited, and at high extinction the optical depth is significantly larger than $1$  \citep{Pineda07}.  We quantify and discuss the uncertainties in molecular column density as compared with dust measurements below in \S \ref{uncertainty}.

\section{Data Editing} \label {editing}

In order to restrict the comparison in this paper to reliable measurements of column density in the portions of molecular clouds {\it not} dominated by the localized heating and stirring caused by embedded massive stars we have excluded certain positions from all three data sets compared here.  The ``exclusion criteria" are as follows.  

Points not having reliable measurements in all three tracers are omitted.  For \co\ and \thco\ data, we use only positions with signal-to-noise (peak-to-rms) ratios greater than 15 and 5, respectively. Note, though, that even with infinite sensitivity there would still be a minimum detectable column density for the line measurements, due to the fact that a critical density of matter is needed to excite CO or \thco\ collisionally.  (This ``minimum" density for excitation is what leads to the additive constant in the relationship between \av\ and $N(\mthco)$ shown in eq..~(\ref{eq:n13-lte}).  We also exclude any points where the fitted \co\ line-width is smaller than 80\% of the \thco\ line-width, because this is indicative of either: 1) multiple velocity components along the line of sight captured by one tracer but not the other; or 2) of an unphysical result caused by marginal data quality.

The IC348 and NGC1333 cluster regions are eliminated, using purely spatial cuts (see black ``holes" in Figure~\ref{f1}), from our analysis, because when material is heated non-uniformly from within, it is difficult to derive accurate column densities from either dust or gas emission \citep[see][]{Schnee06}.  We also remove a small number of other pixels with stellar densities high enough (in this case, $>10$ stars/pixel) to imply a significant contribution from stars embedded in Perseus.  This is necessary because NICER relies on ``background" stars for extinction mapping, and will underestimate extinction if embedded stars are included by accident.

As indicated on the dust temperature map in Figure \ref{f1}d, a heated shell surrounds the
B-star HD 278942, and portions of it overlap with our study region, in
projection.  The shell is actually located just behind the molecular
clouds, and it apparently touches the clouds at a few localized points
of contact \cite[see][]{Ridge06b}.  In 
Figures~\ref{f1} and \ref{f3},  we have marked the effects of the shell's low-density hot interior (in pink), as well as of its over-dense heated rim (in blue). In the histograms of Figure~\ref{f2} and for the corresponding fits shown in Table~\ref{fits}, the points effected by the shell's interior and exterior have been {\it excluded}.  In Figure~\ref{f3}, we show, for illustrative purposes only, the shell interior/exterior points in pink/blue, but we calculate the scatter amongst the different tracers {\it excluding} the shell points.

Lastly, positions close to the border of the map are removed when they would be effected by inaccurate convolution.

In summary, all of the positions shown with non-null values in Figure~\ref{f1} are shown in Figure~\ref{f3}, but the points within the ``shell" regions are excluded from the histograms and fits in Figure~\ref{f2}.

\section{Results: The Distribution of Column Density in Perseus}
\label {results}

\begin{figure}
\epsscale{1.2}
\plotone{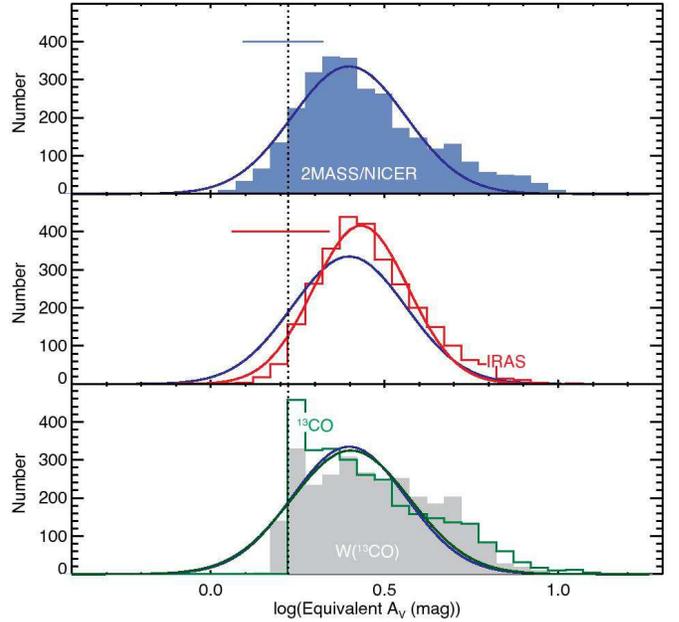}
\caption{Pixel-by-pixel column density histograms for the full Perseus COMPLETE data set shown in Figure \ref{f1}. Note that all tracers are smoothed to a common $5'$ resolution, and only points where \thco\ is reliably detected are included (see text).  Each distribution shows the column density as labeled on the plot, converted to units of \av, as explained in the text.  In every panel, the solid smooth blue curve shows the Gaussian fit to the 2MASS/NICER-based distribution, for reference.  The red smooth curve in the middle panel, and the green smooth curve in the lower panel, show Gaussian fits to the IRAS- and \thco-implied  distributions, respectively.  The grey shaded histogram in the bottom panel shows $W$(\thco) converted to units of \av, but the fit shown is only for the curve just labeled ``\thco," which gives total column density calculated using eq.~(\ref{eq:n13-lte}).   Fit parameters for all four column density distributions shown in this Figure are given in Table \ref{fits}.  The dashed vertical line extending through all three panels shows the cutoff (lowest) value of column density measurable with \thco, according to eq.~(\ref{eq:n13-lte}). The short horizontal bar centered on the dashed line in the top pane shows the $1\sigma$ spread in distribution of  $|\mav(2MASS)-\mav(\mthco)|/\mav(\mthco)$ and for the middle pane the same dispersion but for $|\mav(IRAS)-\mav(\mthco)]/\mav(\mthco)|$.  Figure~\ref{f4} shows the regional breakdown of these same histograms.
\label{f2}}
\end{figure}

Figure \ref{f2} compares, as histograms, the distributions of column density
implied by the three different techniques (extinction, thermal
emission, and molecular lines) under study here.  To facilitate
comparison with theories that predict a log-normal density
distribution (see \S \ref{sim}), the histograms are shown in linear-log
space, where a log-normal appears as a Gaussian.  The log-normal
least-squares fit to the 2MASS/NICER-based extinction distribution is
repeated (in blue) as a fiducial in all plots.  Table~\ref{fits} shows the fit parameters for each column density distribution based on the functional form:

\begin{equation}
{\rm Number\ of\ pixels} ={\rm Scaling}*\,e^{-(\log \mav - \log {A_{V,0}})^2/2\sigma^2}  
\end{equation}

\begin{deluxetable}{lcccc}\label{fits}
\tablecolumns{5}
\tablewidth{0pc}
\tablecaption{Fit Parameters 
\label{fits}}
\tablehead{
\colhead{Tracer }&
\colhead{Scaling}&
\colhead{$\log A_{V,0}$}&
\colhead{$A_{V,0}$}&
\colhead{$\sigma$}
}
\startdata
\av 			& 336$\pm$14	& 0.397$\pm$0.008	&   1.18$\pm$0.01 	& 0.163$\pm$0.008\\
$N_{dust}$	& 416$\pm$12	& 0.433$\pm$0.005	& 1.145$\pm$0.005 	& 0.136$\pm$0.005\\
$N(\mthco)$	& 324$\pm$27	& 0.40$\pm$0.02 	&   1.19$\pm$0.02 	& 0.17$\pm$0.02  \\
$W(\mthco)$	& 298$\pm$18	& 0.45$\pm$0.01 	&   1.22$\pm$0.02 	& 0.20$\pm$0.01
\enddata
\end{deluxetable}

It might seem surprising, especially given the non-gaussian shape of the \thco\ histogram, that the $A_{V,0}$  parameter for all the column density tracers are so similar, however, keep in mind that this is to be expected because thermal emission and molecular line data have been calibrated with the 2MASS/NICER-based extinction map.

\subsection{Uncertainty Based on Tracer Inter-Comparison} \label {uncertainty}

\begin{figure}
\epsscale{1.15}
\plotone{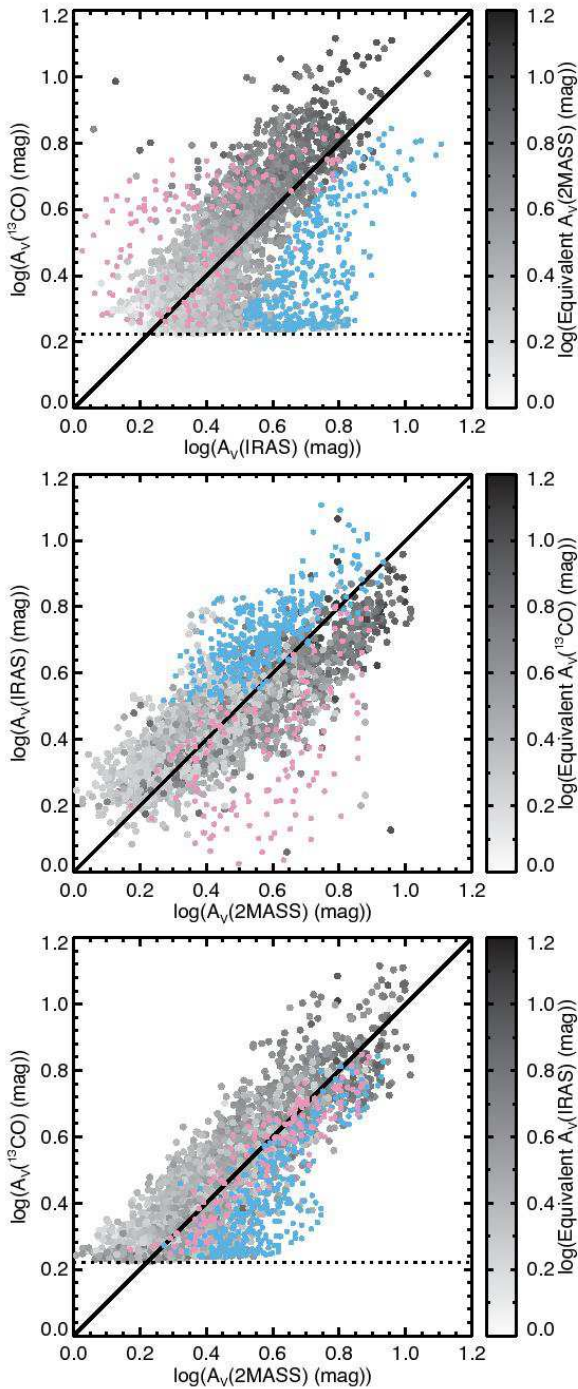}
\caption{Inter-comparison of measured column density distributions, all shown in units of \av~(see text).  The horizontal dotted lines in the top and bottom panels show the minimum column density measurable with \thco\ corresponding to the additive constant in eq.~(\ref{eq:n13-lte}).  Note that the grey scale color of the data points shows the ``third" measure of column density not plotted on the $x$ or $y$ axis for each plot, so that any points that do not look to be part of a smooth grey gradient are ``outliers" in that third measure.  Points colored blue and pink correspond to the blue-outlined shell exterior and the pink-outlined shell interior in Figure~\ref{f1}.  The pink and blue point are shown here for illustrative value, but they are not included in the histograms in Figures~\ref{f1} and \ref{f4}, or in any fits to column density distributions, as explained in the text.  The 45-degree lines are not fits: they simply show a 1:1 relationship that might reasonably be expected given that the IRAS-based and \thco-based measures have been calibrated to best match the 2MASS-based \av\ distribution overall (see text).  
\label{f3}}
\end{figure}

Figure \ref{f3} shows direct inter-comparisons of the all three column density measures: extinction, thermal dust emission, and gas emission.  The overall scatter around the 1:1 lines can likely be attributed, in large part, to variations along the line of sight of: dust properties; dust temperature; and gas volume density (which effects line emission). We explain below that the principal source of variation between the two dust tracers is variation in temperature along the line of sight; while for comparisons between \thco\ and dust tracers, variations are caused by changes in the ratio of gas to dust along the line of sight.  

Dust emission is dependent on temperature, but extinction is not.  By applying radiative transfer calculations to numerical models of molecular clouds, \citet {Schnee06}  showed that most of the scatter in comparing extinction and emission based column density measures is likely caused by line-of-sight variations in dust temperature.  Specifically, the scatter in the middle panel of Figure \ref{f3} is directly modeled, and explained, by this effect in \citet {Schnee06}, so we will not discuss it further here.

\begin{figure}
\centering
\epsscale{1.2}
\plotone{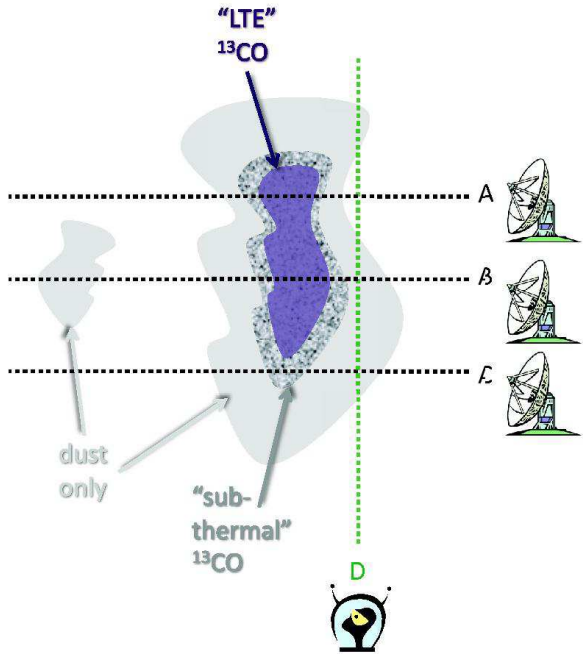}
\caption{Schematic diagram showing various lines of sight through various conditions.  A, B, and C, are views as seen from our vantage point on Earth.  In A, we have the favorable situation where \thco\ is in near-LTE (shaded purple), and all of the low-density material around it, some of which does not emit in \thco\ at all (shaded grey), but still shows up in dust measures, is associated with the cloud of interest.  In B, the situation is as in A, but an additional region which emits in dust but not in \thco\ (either due to low density or due to a dearth of \thco) is included.  In C, the densest gas the line of sight passes through (shown as marbled grey) is dense enough to excite some \thco, but not at a level truly indicative of the full column density present, because the collisions at this ``sub-critical" density are too infrequent.  In D, we see an ``alien's view" of the same cloud, which passes through the ``dust-only" zones but no \thco\ emitting regions, even though it crosses A, B, and C.
 \label{f6}}
\end{figure}

To discuss the variations between dust- and gas-based measures of column density, we use Figure~\ref{f6}, which shows a schematic diagram of gas and/or dust structures viewed from various vantage points.   The general idea of a ``dust-to-gas" ratio is fine as long as one takes care to specify the {\it volume} over which that ratio is intended to apply.  In other words, if clouds like those shown in the Figure~\ref{f6} made up a whole galaxy, then summing all the dust and all the gas mass and dividing would be a fine and accurate expression of a volume-averaged dust-to-gas ratio.    In our case, however, where we are interested in detailed pixel-by-pixel comparisons of column density measured with either dust or gas, plane-of-the-sky variations in the dust-to-gas ratio matter greatly, but are difficult to treat.  

The full ``dust-to-gas" ratio derivable from measurements of  CO isotopes and extinction relies on the wavelength dependence of extinction and on the abundance of CO isotopes in the gas.   As explained in  \S\ref{data-molecular-line} and in \citet{Pineda07}, we assume in this paper that the wavelength dependence of extinction is constant (as $R_V=\mav/E(B-V)=3.1$), and that the abundance of \thco\ is fixed for all of Perseus, at the value corresponding to the coefficients in eq.~(\ref{eq:n13-lte}).  We make these assumptions in order to show how gas and dust maps would {\it typically} compare, as this approach, or an even less customized one, where dust-to-gas ratios and abundances from the literature are used without testing, is most common.  In  \citet{Pineda07} we give a detailed analysis of how much the calibration of dust-to-gas relationships (\thco\ abundance) varies amongst the sub-regions of Perseus under both the linear approximation (eq.~(\ref{eq:n13-lte})) and also with a curve-of-growth approach. 

The \thco (1-0) transition is only excited above a certain critical density, and the transition to ``LTE" levels of excitation is actually a gradual one.   So, in Figure \ref{f6}, we show some low-density dusty regions where \thco\ is not excited at all, some  slightly-higher-density regions where it is sub-thermally excited, and some where the density is high enough for near-LTE conditions.  Figure \ref{f6} clearly shows that while we can excise points with no \thco\ emission on the plane of the sky, we cannot excise dust along the line of sight that would produce no \thco\ emission if viewed ``from the side" (e.g. by the alien in the Figure).

If all lines of sight in a region were exactly the same as line-of-sight ``A" in Figure \ref{f6}, then: 1) an equation for converting gas to dust column density like eq.~(\ref{eq:n13-lte}) could be valid; and 2)  there would be no scatter in comparing dust and gas measures of column density.    However, in a real cloud like Perseus, some lines of sight pass through additional material that contains dust but does not emit in \thco,  as shown for line of sight ``B" in Figure \ref{f6}.  That additional material can be background or foreground to the cloud. Or, it can be material within the cloud that is depleted or deficient in \thco, for a variety of reasons.  

The \thco\ column densities given in this paper are calibrated using a single ``typical" conversion factor in eq.~(\ref{eq:n13-lte}), for all of Perseus.  This calibration makes the implicit assumption that  the mixture of foreground/background/depleted/deficient material along lines of sight within Perseus does not vary too much, which is not strictly true--and the scatter seen in Figures~\ref{f2} and \ref{f4} is caused primarily by this assumption.  Note that since we {\it know} that the shell region (highlighted using blue and pink in Figures~\ref{f1} and \ref{f3}) is unusual in both its temperature and for its apparent \thco\ deficiency (see \S \ref{paucity}, below), it has been excised from all fits (including the overall calibration) and histograms (Figures \ref{f1} and \ref{f4}) in this paper.  

Equation~(\ref{eq:n13-lte}) has an additive factor corresponding to the typical minimum column density at which \thco\ is detected.  This minimum is shown with dotted lines in Figures \ref{f2} and \ref{f4} at  $\log \mav=\log 1.67=0.22$.   Just as the slope of the gas-to-dust calibration does not universally apply to all lines of sight, neither will this minimum.   As explained above, we have excised all positions with \thco\ levels below the minimum set by eq.~(\ref{eq:n13-lte}), but the remaining positions will sometimes give dust-derived extinctions below the gas-derived minimum.   The biggest effect causing ``leakage," beyond the low-density cutoff, which is partially responsible for the low-density tails seen in Figures \ref{f2} and \ref{f4}, is simply the scatter about the 1:1 relationships shown in Figure \ref{f3}.  This scatter is characterized by horizontal error bars in Figures~\ref{f2} and \ref{f4}, the calculation of which is explained below.   

A more subtle effect causing ``leakage" below the \thco\ low column cutoff is also present, and it is illustrated in line of sight ``C" in Figure \ref{f6}, which shows  a line of sight that passes through a ``sub-thermally excited" region.   In cases like ``C," \thco\ will be detectable, but the density in the region producing the emission is below or barely at the critical density, such that one cannot reliably convert \thco\ line intensity to a column density using eq.~(\ref{eq:n13-lte}). As shown in \citet{Pineda07}, when one uses a more appropriate curve-of-growth fit, rather than a linear approximation, \thco\ integrated intensity typically rises {\it more} steeply than a linear fit to the same data would imply.  As a result, a linear conversion (eq.~(\ref{eq:n13-lte})) applied to the observed \thco\ intensity in low-density (sub-thermally-excited) regions will often overestimate the column density.  Thus, some of the positions creating the ``low column" tails shown in Figures \ref{f2} and \ref{f4} are likely to actually be at the low column densities that dust measures: their column density is just overestimated by (sub-thermally-excited)  \thco.  Note, further, that the seemingly odd pile-up of \thco\ at densities just above the cutoff in Figure \ref{f2} is similarly caused by the linear approximation's inability to properly treat sub-thermally-excited (relatively low density) gas.

To empirically estimate the scatter associated with all manner of variations in dust and gas properties, we calculate the standard deviation of the distributions of the normalized differences between one column density measure and another at each point in the maps. We find, for the full Perseus map, that that $|\mav$(IRAS)-\av(2MASS)$|/\mav$(2MASS) has  1-sigma width (standard deviation) of 26\% and $|\mav$(\thco)-\av(2MASS$)|/\mav$(2MASS) has $1 \sigma$ width 24\%.   It is clear from Figure \ref{f3} that these standard deviations will be larger than the $1 \sigma$ uncertainty in a linear fit, because the 1:1 line (implicitly assumed in the standard deviation calculation) is not a perfect representation of the real relationships between all the extinction measures. 

In Figures \ref{f2} and \ref{f4}, we show horizontal bars of varying lengths centered at the threshold value of $\log \mav=0.22$  in order  to demonstrate that the fluctuations one sees about the 1:1 lines in Figure \ref{f3} are large enough to cause the low-column-density tails seen in Figures \ref{f2} and \ref{f4}. The specific length for each horizontal bar shown is the standard deviation of $|\mav$(IRAS)-\av(\thco)$|/\mav$(\thco) for IRAS-based panels of Figures \ref{f2} and \ref{f4}, and  $|\mav$(2MASS)-\av(\thco)$|/\mav$(\thco) for the 2MASS/NICER-based panels.  We discuss the implications of dust and gas probing slightly different regions along the line of sight further in \S \ref{reliability}, below.

\subsection{Paucity of $\mthco$ in the Shell Around HD 278942} \label {paucity}
 
The overabundance of blue points in the lower-right portion of the two parts of Figure~\ref{f3} that involve \thco\ data is likely due to a dearth of \thco\ in the shell around HD 278942.  On the plane of the sky, nearly all of the points that lie below the overall 1:1 trend are associated with the shell (see blue contour in Figure~\ref{f1}).  The effect appears most strongly when comparing \thco\ to dust emission (Fig.~\ref{f3}, top), but we base our statements here on the comparison of \thco\ and extinction (Fig.~\ref{f3}, bottom), because the shell's effects also biases the dust emission column density measurements in that region (as shown in the middle panel of Figure~\ref{f3}).  We suspect that either the shell structure is young, and molecules, such as \thco, have not had time to form, or that the energetic radiation associated with the shell \citep[see][]{Ridge06b} has destroyed pre-existing \thco.  In other words, the column density along lines of sight through the shell (outlined in blue in Figure~\ref{f1}) has a molecular component, associated with the (rest of the) Perseus star-forming region, plus a component associated with the shell which contains a lower fraction of molecular gas.

\subsection{Regional Variations} \label {regvar}

\begin{figure*}
\centering
\epsscale{1.0}
\plotone{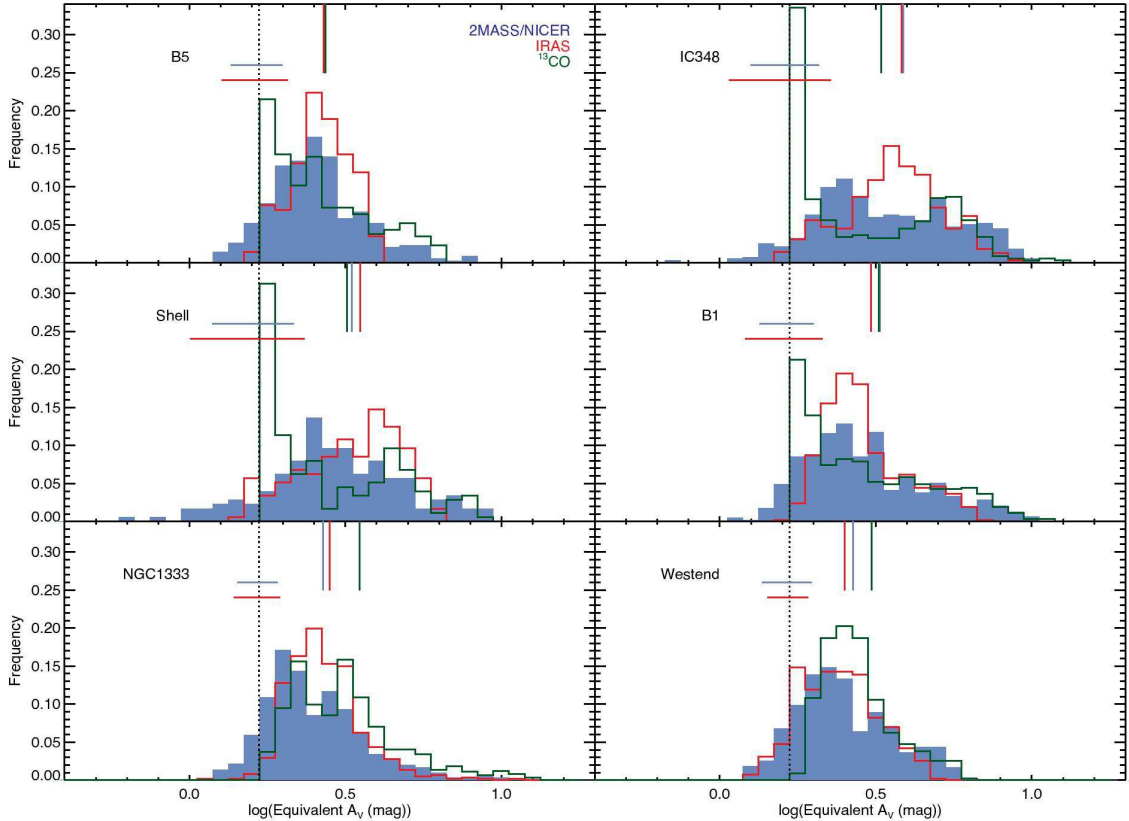}
\caption{Regional variations in column density distributions.  Each panel shows the same assortment of measured distributions as in Figure \ref{f2} only without {\it W}(\thco).  As explained in more detail in Figure \ref{f2}'s caption and in the text, the vertical dotted line shows the minimum \av\ measurable with \thco\ and the short horizontal bars show indicate the characteristic scatter about the relationship between each dust tracer and \thco.  Parameters for Gaussian fits to normalized versions of these distributions are given in Table \ref{regions}.   The short vertical lines hanging from the top axis indicate the mean value of extinction for each distribution (by color), and the means of the normalized extinction are given in Table \ref{regions}.  Note that we use ``frequency"  on the y-axis in these plots, rather than ``number" as in Figure \ref{f2}, to facilitate inter-comparison of the various regions, not all of which include the same total number of positions (see Table \ref{regions}).
\label{f4}}
\end{figure*}

Perseus itself is a large complex of molecular clouds.  A division of it into a set of morphologically-distinct sub-regions can be guided by a plot of dust temperature versus gas velocity, as explained in \citet{Pineda07}.  Histograms of column density analogous to those shown in Figure~\ref{f2} are shown in Figure~\ref{f4} for the same sub-regions (B5, IC348, Shell, B1, NGC1333, and Westend) discussed in  \cite{Pineda07}.  Significant variation from region to region is clearly present.  Typically, each sub-region has of order 500 points (out of $\sim 3500$ for all of Perseus, see Table~\ref{regions}) and has a maximum linear extent of $\sim 5$ pc rather than the $\sim 25$ pc full length of Perseus, making each sub-region a significantly smaller statistical and physical sample than is all of Perseus.

\begin{deluxetable}{lcccc}\label{regional}
\tablecolumns{5}
\tablewidth{0pc}
\tablecaption{Regional Variations
\label{regions}}
\tablehead{
\colhead{Name }&
\colhead{Mean \av}&
\colhead{Mean $\ln{\mav}$}&
\colhead{Mean $\ln{x}$}&
\colhead{Sigma $\ln{x}$}
}
\startdata
\cutinhead{All, Number of Points= 2892}\\
NICER     & 3.080 &  1.024  & -0.101  &   0.433 \\
IRAS      & 3.032 &  1.052  & -0.057  &   0.329 \\
N(\thco)  & 3.243 &  1.076  & -0.101  &   0.427 \\
W(\thco)  & 3.215 &  1.083  & -0.084  &   0.404 \\
\cutinhead{B5, Number of Points=  344}\\
NICER     & 2.734 &  0.940  & -0.065  &   0.349 \\
IRAS      & 2.692 &  0.968  & -0.022  &   0.213 \\
N(\thco)  & 2.850 &  0.975  & -0.073  &   0.368 \\
W(\thco)  & 2.733 &  0.941  & -0.065  &   0.353 \\
\cutinhead{IC348, Number of Points=  552}\\
NICER     & 3.872 &  1.230  & -0.124  &   0.500 \\
IRAS      & 3.826 &  1.279  & -0.063  &   0.359 \\
N(\thco)  & 3.487 &  1.105  & -0.144  &   0.525 \\
W(\thco)  & 3.293 &  1.068  & -0.123  &   0.496 \\
\cutinhead{Shell, Number of Points=  176}\\
NICER     & 3.318 &  1.076  & -0.123  &   0.503 \\
IRAS      & 3.523 &  1.194  & -0.065  &   0.373 \\
N(\thco)  & 3.250 &  1.053  & -0.125  &   0.484 \\
W(\thco)  & 3.193 &  1.053  & -0.108  &   0.458 \\
\cutinhead{B1, Number of Points=  631}\\
NICER     & 3.268 &  1.075  & -0.109  &   0.448 \\
IRAS      & 3.057 &  1.068  & -0.050  &   0.304 \\
N(\thco)  & 3.351 &  1.086  & -0.123  &   0.477 \\
W(\thco)  & 3.233 &  1.069  & -0.104  &   0.450 \\
\cutinhead{NGC1333, Number of Points=  642}\\
NICER     & 2.681 &  0.915  & -0.071  &   0.358 \\
IRAS      & 2.813 &  0.994  & -0.041  &   0.271 \\
N(\thco)  & 3.445 &  1.147  & -0.090  &   0.396 \\
W(\thco)  & 3.518 &  1.190  & -0.068  &   0.356 \\
\cutinhead{Westend, Number of Points=  547}\\
NICER     & 2.674 &  0.919  & -0.065  &   0.353 \\
IRAS      & 2.511 &  0.879  & -0.041  &   0.285 \\
N(\thco)  & 2.879 &  1.021  & -0.036  &   0.259 \\
W(\thco)  & 3.071 &  1.089  & -0.032  &   0.250
\enddata
\end{deluxetable}

The shapes of the distributions vary significantly from sub-region to sub-region, and most also show significant disagreement amongst the three column density tracers used here.  One key point is that some of the tracer-to-tracer disagreement is caused by (purposely) using only a single CO abundance, and a single form of eq.~(\ref{eq:n13-lte}) in creating Figure \ref{f4}.  Had we customized the calibration region-by-region, as is done in \citet{Pineda07}, we could improve the tracer-to-tracer agreement some, but then we would not be representing legitimate sub-sets of the same data shown in Figure \ref{f1} through \ref{f3}.  Overall, \thco\ as a tracer seems most capricious, and we suspect that this is due to significant variations in physical conditions (e.g. temperature, radiation field, region age) other than (column) density that effect its abundance and excitation.

A subtler point concerning variations in distribution shape, both regionally, and tracer-to-tracer, concerns real variations in dust-to-gas ratio, in gas properties (e.g. \thco\ abundance), and/or in dust properties (e.g. $R_V$).  We know that these properties vary \citep[cf. \S
\ref{data-molecular-line} and ][]{Pineda07} even on the many-pc scales characteristic of the ``regions" we discuss here.  So, they likely also vary on smaller scales, meaning that subtle changes in the shape of the distributions seen in Figure~\ref{f2} and \ref{f4} would be apparent if we could account for these variations.  In their paper on ``Can We Trust the Dust?," \citet[][see also \S\ref{reliability}, below]{padoan06} suggest that very small-scale variations ($<0.1$ pc) variations in the dust-to-gas ratio may in fact exist.  So, in the future, when we can carry out inter-comparisons of column density tracers on even smaller scales than those we consider in this paper, it will be interesting--and potentially important--to quantify how much the dust-to-gas ratio, as well as intrinsic gas and dust properties, change at various scales, and under various conditions.

If we study only the NICER-based histograms in Figure~\ref{f4}, which are not effected by calibration choices here, there are still very significant variations from region to region, and one is left wondering how sample (region) size effects the shape of one of these histograms.   We consider this question, and others, in the context of numerical simulations, below.

\section{Implications}
\subsection{Comparisons with Numerical Simulations}
\label{sim}
In this section, we consider how numerical simulations offer insight into the shape of column density distributions.

\citet{v-s94} shows that for highly supersonic flows where gravitational and 
magnetic forces become negligible, the gas has a pressureless behavior. 
Under these conditions, the hydrodynamic equations become scale invariant, i.e., motions at
all length and density scales obey the same equations. 
As a result, the probability density function of the volume density, $n$, is expected to be log-normal. \citet{ostriker} demonstrate that for essentially isotropic flows, the same kind of log-normal distribution results for either volume or for column density.  These general predictions of turbulence theory inspire the log-normal fits shown in Figure \ref{f2}. We do not suggest that any of the column density distributions we find here are necessarily best-fit by log-normals--we simply offer log-normal fits as a relevant comparison.

Figures \ref{f1} and \ref{f4} straightforwardly show number (Figure~\ref{f1}) and frequency (Figure~\ref{f4}) distributions of $\log N$.  If, however, one wants to emphasize density {\it fluctuations} about a mean, it makes sense to normalize the distributions by a mean column density.  Most numerical simulations, because they are often scale invariant, analyze this kind of ``normalized column density," so to facilitate comparison, we define
\begin{equation}\label{eq:x}
x=N/\bar N 
\end{equation}
where $\bar N$ is the mean value of column density in any map.  The values of such means are given in Table \ref{regions} and shown as vertical long ticks in Figure \ref{f4} for all the data sets considered here.  

We define the standard deviation of the distribution of $\ln x$ as $\sigma_{\ln x}$, which would equal the $1\sigma$ standard deviation of a Gaussian fit to the distribution of $\ln x$ in the case of a pure log-normal column density distribution.  The values listed for $\sigma_{\ln x}$ in Table \ref{regions} are determined purely from the statistics of the distribution of $\ln x$ values, though, and {\it not} from any kind of (e.g. Gaussian) fit.

Some simulations have suggested that there may be measurable relationships amongst the mean and dispersion of the distribution of $\ln x$, Mach number, mean magnetic field strength, and the ratio of forcing scale to cloud scale.  For example, \citet{padoan97b} suggest that
\begin{equation}
{\sigma_{\ln x}}^2=\ln(1+  \mathcal{M}^2\beta^2)
\label{paolobeta}
\end{equation}
where $\beta$ is a constant of order 0.5 and $\mathcal{M}$ is the sonic Mach number of the gas.   However, if a magnetic field is present, both \citet {padoan97b} and \citet{ostriker} find that this relationship changes.   In particular, \citet{ostriker} find a secular trend in density contrast that depends on the fast magetosonic Mach number, $\mathcal{M}_F$, which in turn depends on the sound speed and the Alfv\'en speed.  Presently, region-by-region detailed measurements of field strength, needed to measure the Alfv\'en and sound speed independently, are not available for any set of regions as large as those we study here, so it is hard to test this relationship directly with observations right now.   However, a concerted effort to test predictions that rely on $\mathcal{M}_F$, perhaps using the Chadrasekhar-Fermi method to estimate field strengths over large areas, could and should be undertaken in the near future.

Even if we could know field strengths though, it is still not the case, according to simulations, that a single realization of a turbulent flow can be inverted to give basic physical parameters, such as the exact power spectrum of density fluctuations in the flow.  Several researchers \citep[e.g.][] {v-s94, ostriker} have shown that sub-samples of a single flow and/or multiple realizations of the same physical conditions will give noticeably different (normalized) column density distributions, similar to what is seen in Figure \ref{f4}.  There are a large number of modes present simultaneously in any simulation so one sample alone is unlikely to sample them all in their ``average" mixture.  

So, given that simulations demonstrate that a large(r) sample of realizations would be needed to causally relate basic physical parameters to column density distributions, what can we learn from just the data we present in this paper about the physics of turbulent flows in star forming regions?

We can consider our set of ``regional" sub-samples as multiple realizations of the same ``experiment" we can call the ``Perseus Molecular Cloud."    In one sub-sample, the ``Shell" region, where the gas is dominated by an obvious driver not present in the other regions, we might dismiss any outlier-like behavior as due to ``unusual" forcing, and there is in fact a skewing of its column density distribution toward lower values, not seen in the other regions (see Figure \ref{f4}).  

\begin{figure}
\centering
\epsscale{1.2}
\plotone{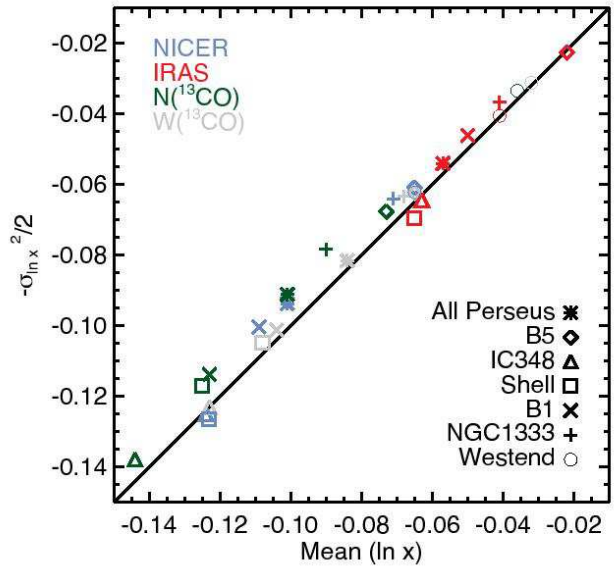}
\caption{Comparison of the fit parameters listed in Table \ref{regions} with each other, and with the 1:1 line predicted, for a log-normal, by eq.~(\ref{disp}).
\label{f5}}
\end{figure}

If the regional density distributions are truly drawn from an underlying distribution which is inherently log-normal, as predicted by many simulations, then 
\begin{equation}
{\overline{\ln x}}=-{{\sigma_{ln x}^2}\over 2}
\label {disp}
\end{equation}
simply because the mean (first moment) of a log-normal distribution of a quantity (e.g. $x$) that is {\it normalized} by its own mean should be unity.  Thus, Figure \ref{f5}, which shows $-{\sigma_{\ln x}^2/2}$ as a function of the mean $\ln x$ for each region, appears to indicate that nearly all of the distributions we study in this paper are close to consistency with being drawn from a log-normal. Note, however, that distributions similar in shape to a log-normal will also give a relationship very similar to eq.~(\ref{disp}), so the fact that the points for our regions lie so close to the line should only be taken to mean that the distributions are close to consistent log-normal, and not exactly log-normal.  The main purpose of Figure \ref{f5} is to show all the values shown in Table \ref{regions} together, in a way that facilitates testing hypotheses relying on making comparisons.

If a relationship similar to eq.~(\ref{paolobeta}) were satisfied, then moving up and right along the ``perfect log-normal" line in Figure \ref{f5} would mean higher Mach number in the hydrodynamic case, or possibly higher fast magnetosonic Mach number (as shown by \citealt{ostriker}, in their Figure 4).  Note that  $\mathcal{M}$ will rise if the sound speed drops, and  $\mathcal{M}_F$ will rise {\it either} if the magnetic field drops or if the sound speed drops, for a given density of material.  We have tested whether a sonic Mach number determined from observed line widths and (dust) temperatures listed in \citet {Pineda07} increases along the 1:1 line in Figure \ref{f5}, and it does not.   

Making comparisons based on Mach numbers calculated using only {\it dust} (not gas) temperature and {\it line-of-sight} velocity dispersion limits direct observation-simulation comparison in two important ways.  First, we simply cannot calculate $\mathcal{M}_F$ \citep[as is used in][]{ostriker}, which is likely more relevant than $\mathcal {M}$ in comparisons, because we do not have field strength measurements.  Second, in observations, the observed column distribution arises from plane-of-sky compressions, whereas the line widths used to compute Mach number come from the line-of-sight motions only.  We consider how one might get around these limitations in the future in \S \ref{future} below.

One important point to take away from observation-simulation comparisons concerns the sometimes-overlooked distinction between volume density and column density distributions. Several simulators have pointed out that the {\it column} density distributions we show here will only look the same as the {\it volume} density distributions, which are usually predicted to be log-normal, when turbulence is dominated by large-scale motions \citep[e.g.][]{ostriker} .   The fact that the distributions look as close to log-normal as they do suggests that turbulence in Perseus\footnote{Note that we have excised regions from ``Perseus" that are likely to contain, and be effected by, the most local sources of turbulence: the densest portions of the NGC1333 and IC348 young clusters have been excluded from our analysis, so the stellar wind and outflow drivers they contain are not included in the area analyzed here.}  is primarily driven by motions on scales nearly the size of the whole cloud, or larger.  

\subsection{How common are log-normal-like density distributions?}

A similar relationship to the one shown in Figure \ref{f5}, between the mean and dispersion in column density, has been used to explain and interpret the results of \citet{Lada:IC5146}, who we believe were the first to find a relation between the standard deviation in measured extinction and the extinction value itself. \citet {Thoraval} pointed out that a relation like eq.~(\ref{disp}) can be caused by small scale structure below the resolution limit of an extinction map. \cite{padoan97} performed a reanalysis of Lada et al.'s IC5146 observations, and find that the $\sigma _{\mav} - \mav$ relation is consistent with a log-normal column density distribution producing unresolved density structure.  Later, \cite{ostriker} used higher resolution observation in IC5146 by \cite{lal99} to compare the  cumulative column density distribution with their own simulations finding that they are very similar in shape.

Other extinction studies might provide notable exceptions to the ``log-normal" trend.  For example, the Pipe Nebula's  column-density distribution (determined using the 2MASS/NICER method) displays a complex shape with multiple peaks \citep{lom06}. Lombardi et al. attribute these peaks to the possibility of background clouds observed in projection with (and thus in addition to) the Pipe at low Galactic latitudes.  Recent C$^{18}$O observations in the Pipe \citep{Muench07} have shown that there are two distinct velocity components there, and so it is possible that if those distributions could be separated, each would have a more log-normal-like column density distribution.

In general, it is important to consider what boundaries, in either spatial or velocity dimensions, are placed upon regions when assessing how common log-normal distributions might be.  Overlapping ``clouds" along the line of sight may appear as one structure on the plane of the sky, and, without kinematic or distance information, it is hard to separate these structures.  And, even if one can separate populations along a line of sight, the meaning of the region one considers on the plane of the sky needs to be understood as physically ``small" or ``large" in the context of a particular question, such as how ``log-normal" a distribution might be expected to look.

\subsection{Which Tracer Tells us What, how Reliably?} \label {reliability}

Recently, in a paper entitled ``Can We Trust the Dust?", \citet{padoan06} have raised questions about the reliability 
of NIR extinction mapping, based on their finding that the power spectrum of a
NIR extinction map of the Taurus region is significantly shallower
than that derived from a \thco\ map of the same region. Through
detailed modelling, they rule out depletion of CO, and/or CO formation
timescales as the cause(s) of this discrepancy. By producing synthetic
NIR-extinction observations of simulations, they show that the power
spectrum derived from observations should trace the power
spectrum of the actual spatial distribution of dust if the extinction
is proportional to dust column density. 

 \citet{padoan06} argue  that a discrepancy 
between the power spectra of the dust and the gas might be expected in the case of transonic, nonmagnetized turbulence. Using \thco\ and NIR data as we do here, but for Taurus, they interpret a shallower power spectrum in dust fluctuations than in gas fluctuations as evidence for
intrinsic spatial fluctuations in the dust-to-gas ratio, with amplitude increasing toward smaller scales.  Strictly, the  \cite{padoan06} work is only applicable at very small scales (their estimate is $<0.1$ pc).  Thus, the apparent disagreement between our results, which show the gas to be less reliable, and theirs which question the veracity of dust as a tracer of column density, may be only a question of scale.

On the other hand, as shown by \cite{Pineda07} in Perseus, the \thco-dervied column density estimates are adversely effected by 
optical depth even at $\mav\sim4$~mag, and variation in the \thco\ 
abundance with respect to \htwo\ is found between regions in the 
same molecular cloud. These caveats in the interpretation of the \thco\ 
emission maps are not fully taken into account in the \cite{padoan06} analysis (which assumes \thco\ to trace density faithfully out to  $\mav\sim10$~mag), and may also explain the apparent disagreements--both between gas and dust, and between our results and ``Can We Trust the Dust?."

 We have clearly demonstrated, in every figure in this paper, that gas-based and dust-based measures of column density almost never agree perfectly in detail.  So, instead of ``Can We Trust the Dust?", one might ask instead ``Should we Sass the Gas?".  As we discussed in \S \ref{uncertainty}, dust and gas measures are virtually never tracing exactly the same portions of any particular line of sight, so some amount of ``disagreement" when inter-comparing tracers has to be expected.  It also has to be recognized, though, that any particular molecular transition can only trace a (sometimes very) limited range of volume densities, which is effectively bounded at bottom by the critical density and at top by depletion and/or opacity.   Because it is hard to know the real density range to which a line is sensitive, and to model how linear the conversion from line flux to density is likely to be, gas really does deserve some degree of sassing if one wants to tease its ability to trace column density cleanly.  
 
Dust, however, is not perfect either.  For extinction-based measures, variations in the compositional and size distribution of grains can effect the reddening law.  And, for emission-based measures, variations in grain temperature along the line of sight are impossible to correct for and so impose unavoidable uncertainty \citep {Schnee06}.
 
The principle advantage of Dust over Gas is dynamic range.  No single observation of a gas tracer can sample more than about a factor of a about ten in column density, but, dust-based measurements can span a dynamic range of 50 or (much) more.    For example, in the famous B68 Globule, observations of {\it extinction} span the range $\sim 0$ to 27 mag \citep {B68} while C$^{18}$O ranges from 0.2 to 0.8 K \kms\ and then depletes, and N$_2$H$^+$, which only is detected above about \av\ of 10, spans the range 0.3 to 2.1 K \kms\ \citep[][Figure 1] {Bergin2002}.  In TMC-1C, thermal dust {\it emission} allows for a dynamic range of nearly 25 in \av\  \citep[from 4 to 90 mag, ][]{TMC1Cdust}, but none of the many lines mapped there has a dynamic range of more than ten \citep{TMC1Cgas}.   

Extinction measures will always saturate at some high value of extinction where background sources cannot be detected, and emission is often limited in utility at very low extinctions, due to low flux levels and/or observational strategies (e.g. chopping) that make mapping extended emission difficult.  In cases where one can can cross-calibrate extinction and emission using maps of regions where they should give the same information, one can stretch the dynamic range of ``dust" to its fullest by using extinction as the sole probe at low column densities,  emission alone at the higher ones, and both extinction and emission in the intermediate regime. 

In the maps of Perseus in this paper, we have excised points with no \thco\ emission, but our dust measures still can trace low-density material along those lines of sight that {\it do} emit in \thco\ (see \S\ref{uncertainty} and Figure~\ref{f6}).  In Figure 6 of \citet{Schnee05} we show the dust-based column density distributions for all of Perseus, without limiting ourselves to \thco-emitting regions.  Those distributions look similar to the ones in Figure~\ref{f2}, except that they show a longer high-density tail, caused in large part by {\it not} excluding all of the high-density cluster regions NGC1333 and IC348.  It is very important to realize that our excision of plane-of-the sky points with no \thco\ here does {\it not} exclude low-column regions traced by dust along the line of sight, which is why the distributions in Figures~\ref{f2} and \ref{f4} look more symmetric (not cut-off at the low end) for the dust measures than for \thco.  

The only advantage of Gas over Dust {\it when mapping column density} is the ability it offers to kinematically separate regions along the line of sight.  Otherwise, Dust wins.

\subsection{How to Map Column Density Distributions in the Future?}\label {future}

From the discussion above, we may sound ready to recommend dust as the single best tracer of column density in interstellar space--but we are not.  We cannot forget that the total (gas + dust) column density in any region can only be as accurate as the gas-to-dust conversion factor used. And, since all interstellar regions contain much more gas than dust, a small error in a dust-dervied column will be compounded when converted to total column.

Therefore, we recommend a ``holistic" approach to measuring column density.  We need to take account of the level of uncertainty inherent in each conversion factor (e.g. abundance ratio, reddening-to-extinction ratio, dust opacity, dust-to-gas ratio) and assumption (e.g. about line of sight structure) we use in a column density calculation, and then we need to choose the best solution for each particular question.  For example, in a circumstance where we are relatively confident that the dust-to-gas ratio is not varying, and that line-of-sight blending is not causing confusion, then extinction maps will provide the best handle on the kind of relative column density measurements needed to measure key distribution functions, such as ``clump" mass functions.  On the other hand, in a region where the dust-to-gas ratio is very uncertain (e.g. near an HII region) using molecular line ratios to measure volume density and opacity and converting to column density may be superior.  In other words, we recommend that more care than has been typical in the past be given to making the {\it most appropriate} choice in any particular study.

Lastly, several recent studies have shown that ``observing" simulations using radiative transfer and/or chemistry codes, and synthetic telescopes that mimic biases imposed by real ones (an approach we have called ``Taste-Testing") can sometimes uncover hidden limitations and biases associated with certain techniques \citep [e.g.][]{Schnee06, padoan06}.  Thus, if a realistic simulation can be observed synthetically with various column density probes, it can provide a good guide to which kind(s) of column density measures might be least biased under relevant conditions.

\subsection{Conclusions}\label{conclusions}

Careful re-calibration and inter-comparison of extinction, thermal emission, and molecular emission maps of Perseus has allowed us to conclude that:

\begin{itemize}
\item The column density distribution of material in the full Perseus star-forming region, with $1<\mav<12$ mag, is roughly log-normal, when it is not directly effected by embedded clusters or young stellar outlows (bipolar or spherical). 
\item Dust is superior to molecular lines for tracing out the ``full" mass distribution over the range of extinction studied, because it does not require a threshold density to ``excite" and it does not die out at high densities due to high opacity or chemical depletion, the way \thco\ does.
\item The dearth of molecular gas (\thco) in the region corresponding the shell created by the B-star HD 278942 suggests that either CO has not yet formed in this young structure, and/or that existing molecular gas has been dissociated by the shell's interaction with the cloud.
\item When Perseus is dissected into smaller ``sub-regions," the column density distributions become more ragged, as is predicted by simulations for samples that are statistically small.  However, we find that the sub-regions distributions are still log-normal-like, in that the relationship between their normalized means and their dispersions follows a trend consistent with log-normal distribution.
\item In comparing observations of column density (or mass) distributions with each other, and/or with simulations, it is perhaps more important than has been previously appreciated to account for the effects of biases due to dust temperature variations, abundance variations, opacity effects, and observing strategies.
\end {itemize}

We recommend, for the near-term future, the assemblage of an ensemble of maps of column density, along with measurements of Mach number and magnetic field strength, in order to assess the turbulent properties of molecular clouds and star-forming regions more generally.  This large sample is needed to allow for legitimate comparisons between regions, and of observations with simulations, because any one observation of even the same turbulent flow is not enough to characterize the flow's statistical nature.  With such a large sample, we could carry out much more discriminating comparative analyses than the kind represented by
Figure \ref{f5}, which focuses only on testing whether the conditions relating the mean (first moment) of a distribution to its width (second moment) are correct.  We could, for example, begin to investigate the skewess of these distributions, and to investigate alternative functional forms, which may in fact not be exactly log-normal.  With this larger sample, we could also study the effects of star-formation as a driver of the turbulence, and again compare with simulations of this process.  

The combination of: 1) extensive recent improvements in extinction mapping made possible by large-scale near-infrared surveys; 2) the advent of huge molecular line surveys of relatively high-density tracers; and 3) increases in polarization mapping speed (which leads to Chandrasekhar-Fermi-based field estimates),  should very soon allow for studies large enough to test predictive theories and simulations of molecular cloud topology using unprecedentedly large observational statistical samples, and we look forward to it.

\acknowledgements
This paper was originally inspired by a workshop at the Aspen Center for Physics
in the Summer of 2004 on ``Star Formation in Galaxies," where it seemed that the assembled audience of experts could not agree on the least biased way to measure the ``initial conditions" for stars to
form from molecular gas.  A conversation with Eve Ostriker at that
meeting, about how observations and simulations of star-forming
molecular gas might best be compared, was particularly important.   The quest to offer the most bias- and error-free column density distributions we could publish based on the COMPLETE data in this paper took nearly four years, and it spawned several other papers by our group \citep  {Schnee06, Schnee2007, Ridge06a, Pineda07, Foster08}  The ``2007" version of these distributions and their implications were discussed intensively at the KITP Santa Barbara Workshop on ``Star Formation Near and Far," and we deeply thank Eve Ostriker, Paolo Padoan and Enrique Vazquez-Semadeni for their comments both at and since that meeting.  We
thank Jo\~ao Alves, Michelle Borkin, Paola Caselli, Jonathan Foster, Jens Kauffmann, Di Li,  Marco
Lombardi, and Naomi Ridge for their important contributions to the data and results presented in this work.  
We also are grateful to the anonymous referee for comments that contributed to improvements in this paper.
This material is based 
upon work supported by the National Science Foundation under Grant 
No. AST-0407172.  JEP is supported by the National Science Foundation
through grant \#AF002 from the Association of Universities for
Research in Astronomy, Inc., under NSF cooperative agreement
AST-9613615 and by Fundaci\'on Andes under project No. C-13442. 
Scott Schnee acknowledges support from the Owens Valley Radio Observatory, which is supported by the National Science Foundation through grant AST 05-40399.


\begin{thebibliography}

\bibitem[Alves et al.(2001)]{B68} Alves, J.~F., Lada, 
C.~J., \& Lada, E.~A.\ 2001, \nat, 409, 159 

\bibitem[Arce \& Goodman(2001)]{Arce01} Arce, H.~G., \& Goodman,
        A.~A.\ 2001, \apjl, 551, L171

\bibitem[Bachiller \& Cernicharo(1986)]{bc86} Bachiller, R., 
\& Cernicharo, J.\ 1986, \aap, 166, 283 

\bibitem[{{Barnard}(1927)}]{barnard}
{Barnard}, E. 1927, {A Photographic Atlas of Selected Regions of the Milky Way}, ed. E.~{Frost} \& M.~{Calvert} ({Carnegie Institute of Washington})

  
\bibitem[Bergin et al.(2002)]{Bergin2002} Bergin, E.~A., Alves, 
J., Huard, T., \& Lada, C.~J.\ 2002, \apjl, 570, L101 

\bibitem[{{Bohlin} {et~al.}(1978){Bohlin}, {Savage}, \& {Drake}}]{bohlin:1978}
{Bohlin}, R.~C., {Savage}, B.~D., \& {Drake}, J.~F. 1978, \apj, 224, 132

\bibitem[Cambr{\'e}sy(1999)]{Cambresy99} Cambr{\'e}sy, L.\ 1999, 
\aap, 345, 965 

\bibitem[Caselli et al.(1999)]{caselli99} Caselli, P., Walmsley, 
C.~M., Tafalla, M., Dore, L., \& Myers, P.~C.\ 1999, \apjl, 523, L165 

\bibitem[Cernicharo \& Bachiller(1984)]{cern84} 
Cernicharo, J., \& Bachiller, R.\ 1984, \aaps, 58, 327 

\bibitem[Cernicharo et al.(1985)]{cern85} Cernicharo, J., 
Bachiller, R., \& Duvert, G.\ 1985, \aap, 149, 273 

\bibitem[Draine (2003)]{Draine:Review} {Draine}, B.~T.\ 2003, \araa,
41, 241

\bibitem[Enoch et al.(2006)]{Enoch06} Enoch, M.~L., et al.\ 
2006, \apj, 638, 293 

\bibitem[Evans et al.(2003)]{Evans:C2D} Evans, N.~J., II, et al.\ 
2003, \pasp, 115, 965

\bibitem[Foster et al.(2008)]{Foster08} Foster, J.~B., 
Rom{\'a}n-Z{\'u}{\~n}iga, C.~G., Goodman, A.~A., Lada, E.~A., 
\& Alves, J.\ 2008, \apj, 674, 831

\bibitem[Goodman et al.(1995)]{Goodman95} Goodman, A.~A., Jones, 
T.~J., Lada, E.~A., \& Myers, P.~C.\ 1995, \apj, 448, 748

\bibitem [Goodman et al. (2008)] {dendro} Goodman, A., Rosolowsky, E., Borkin, M., Foster, J., Halle, M., Kauffmann, J., \& Pineda, J. E. 2008, \nat, submitted

\bibitem[Hirota et al.(2008)]{Hirota:2008} Hirota, T., et al.\ 2008, \pasj, 60, 37 

\bibitem[J{\o}rgensen et al.(2006)]{Joergensen06} J{\o}rgensen, 
J.~K., et al.\ 2006, \apj, 645, 1246 

\bibitem[Kirk et al.(2006)]{Kirk06} Kirk, H., Johnstone, D., 
\& Di Francesco, J.\ 2006, \apj, 646, 1009 

\bibitem[Lada et al.(1994)]{Lada:IC5146} Lada, C.~J., Lada, E.~A., 
Clemens, D.~P., \& Bally, J.\ 1994, \apj, 429, 694

\bibitem[Lada et al.(1999)]{lal99} Lada, C.~J., Alves, J., \& 
Lada, E.~A.\ 1999, \apj, 512, 250 

\bibitem[Langer et al.(1989)]{Langer89} Langer, W.~D., Wilson, 
R.~W., Goldsmith, P.~F., \& Beichman, C.~A.\ 1989, \apj, 337, 355 

\bibitem[Lombardi \& Alves(2001)]{Lombardi01} Lombardi, M., \& 
Alves, J.\ 2001, \aap, 377, 1023 

\bibitem[Lombardi et al.(2006)]{lom06} Lombardi, M., Alves, 
J., \& Lada, C.~J.\ 2006, \aap, 454, 781 

\bibitem[Miville-Desch{\^e}nes \& Lagache(2005)]{Miville05}
        Miville-Desch{\^e}nes, M.-A., \& Lagache, G.\ 2005, \apjs,
        157, 302
        
\bibitem[Muench et al.(2007)]{Muench07} Muench, A.~A., Lada, 
C.~J., Rathborne, J.~M., Alves, J.~F., \& Lombardi, M.\ 2007, \apj, 671, 1820 

\bibitem[Nordlund \& Padoan(1999)]{NP:99} Nordlund, 
{\AA}.~K., \& Padoan, P.\ 1999, Interstellar Turbulence, 218 

\bibitem[Onishi et al.(1999)]{onishi} Onishi, T., et al.\ 
1999, \pasj, 51, 871 

\bibitem[Ostriker et al.(2001)]{ostriker} Ostriker, E.~C., 
Stone, J.~M., \& Gammie, C.~F.\ 2001, \apj, 546, 980 

\bibitem[Padoan et al.(1997a)]{padoan97} Padoan, P., Jones, 
B.~J.~T., \& Nordlund, A.~P.\ 1997, \apj, 474, 730 

\bibitem[Padoan et al.(1997b)]{padoan97b} Padoan, P., Nordlund, 
A., \& Jones, B.~J.~T.\ 1997, \mnras, 288, 145 

\bibitem[Padoan et al.(1999)]{Padoan:Bell} Padoan, P., Bally, J., 
Billawala, Y., Juvela, M., \& Nordlund, {\AA}.\ 1999, \apj, 525, 318 

\bibitem[Padoan et al.(2006)]{padoan06} Padoan, P., 
Cambr{\'e}sy, L., Juvela, M., Kritsuk, A., Langer, W.~D., \& Norman, M.~L.\ 
2006, \apj, 649, 807 

\bibitem[Pineda et al.(2008)]{Pineda07} Pineda, J.~E.., Caselli, P. \& Goodman,
        A.~A.\ 2008, \apj, 679, 481

\bibitem[Ridge et al.(2006a)]{Ridge06a} Ridge, N.~A., et al.\ 
2006, \aj, 131, 2921

\bibitem[Ridge et al.(2006b)]{Ridge06b} Ridge, N.~A., Schnee, 
S.~L., Goodman, A.~A., \& Foster, J.~B.\ 2006, \apj, 643, 932 

\bibitem[Rosolowsky et al.(2008)]{Erik:2007} Rosolowsky, E., 
Pineda, J.~E.., Kauffmann, J., \& Goodman, A.~A.\ 2008, \apj, 679, 1338

\bibitem[Schlegel et al.(1998)]{Schlegel98} Schlegel, D.~J., 
Finkbeiner, D.~P., \& Davis, M.\ 1998, \apj, 500, 525

\bibitem[Schnee et al.(2005)]{Schnee05} Schnee, S.~L., Ridge, N.~A.,
        Goodman, A.~A., \& Li, J.~G.\ 2005, \apj, 634, 442

\bibitem[Schnee et al.(2006)]{Schnee06} Schnee, S., Bethell, T., 
\& Goodman, A.\ 2006, \apjl, 640, L47 

\bibitem[Schnee et al.(2007a)]{TMC1Cgas} Schnee, S., Caselli, P., 
Goodman, A., Arce, H.~G., Ballesteros-Paredes, J., 
\& Kuchibhotla, K.\ 2007, \apj, 671, 1839 

\bibitem[Schnee et al.(2007b)]{TMC1Cdust} Schnee, S., Kauffmann, 
J., Goodman, A., \& Bertoldi, F.\ 2007, \apj, 657, 838 

\bibitem[Schnee et al.(2008)]{Schnee2007} Schnee, S., Li, J.,
Goodman, A.~A., \& Sargent, A.~I.\ 2008, \apj, 684, 1228

\bibitem[Thoraval et al.(1997)]{Thoraval} Thoraval, S., Boisse, P., \& Duvert, G.\ 1997, \aap, 319, 948 

\bibitem[V{\'a}zquez-Semadeni(1994)]{v-s94} V{\'a}zquez-Semadeni, E.\ 1994, \apj, 423, 681 

\end{thebibliography}
\end{document}